# Mixed system $Cs_3Cu_3Cl_{8-x}Br_xOH$ with weakly connected Cu-triangles

*Natalija van Well*[*,1,2,3]*, Michael Bolte*[4]*, Claudio Eisele*[2]*, Lukas Keller*[3]*, Jürg Schefer*[3]*, Sander van Smaalen*[2]

[1] Department of Earth and Environmental Sciences, Crystallography Section, Ludwig-Maximilians-University Munich, D-80333 Munich, Germany

[2] Laboratory of Crystallography, University of Bayreuth, D-95447 Bayreuth, Germany

[3] Laboratory for Neutron Scattering and Imaging, Paul Scherrer Institute, CH-5232 Villigen, Switzerland

[4] Institut für Anorganische und Analytische Chemie, Goethe-Universität Frankfurt, D-60438 Frankfurt am Main, Germany

ABSTRACT

To study the relationship between the properties of low-dimensional spin systems with weakly coupled Cu-triangles and their crystal structure, single crystals of $Cs_3Cu_3Cl_8OH$ (1) and the new $Cs_3Cu_3Cl_{7.6}Br_{0.4}OH$ (2) were grown. Both compounds are isostructural and crystallize in a monoclinic structure with space group $P2_1/c$. The magnetic susceptibility of (1) shows a maximum at 2.23 K and of (2) at 2.70 K, which are attributed to antiferromagnetic phase transitions. Furthermore, the magnetization along the b-axis at 1.9 K for both compounds shows a spin-flop transition into a new antiferromagnetic phase. This transition occurs at 0.61 T for (1) and at 2.0 T for (2). The antiferromagnetic order can be suppressed by a magnetic field $B_{C1}$= 1.1 T for (1) and $B_{C2}$= 1.2 T for (2). First single crystal neutron diffraction measured on (1) at different temperatures reveals the magnetic signal on the top of the nuclear reflection at (-1 0 0). Its magnetic ordering temperature was found to be at $T_{N1}$= 2.12(3) K.

I. INTRODUCTION

Low-dimensional quantum spin systems have been a subject of intensive studies in the condensed matter community in the last decades. The effective low dimensionality arises from the directional nature of the chemical bonds. In such systems, the classical ferromagnetic (FM) or antiferromagnetic (AFM) ground states can be suppressed by quantum fluctuations[1-6]. Low-dimensional spin-trimer systems are rare in nature and hence, the magnetic properties of such systems are less explored. Only few examples of purely inorganic compounds with triangles are found, which are arranged in one-dimensional (1D) or two dimensional (2D) magnetic lattices. The layered compound $La_4Cu_3MoO_{12}$ with Cu-triangles in the *ab*-plane, being weakly coupled, show frustrated AFM intratrimer interactions. This compound is an example of geometrical frustration in the quantum limit[7,8]. The magnetic structure in this case is very sensitive to intratrimer and intertrimer interactions. An example for a three-leg ladder is $Sr_2Cu_3O_5$. In this compound, the trimers on the rungs interact strongly with each other, so that an analogy to an $S = 1/2$ AFM chain is confirmed[9]. In $A_3Cu_3(PO_4)_4$ (with A = Ca, Sr, Pb), the trimers are arranged in a linear chain, where the intertrimer AFM/FM exchange is smaller than the intratrimer AFM exchange[10-12]. The recently discovered compound $Cs_3Cu_3Cl_8OH$ is a model material for studying the magnetic structure of trimer units with weakly connected Cu-triangles[13]. In spin systems with a triangular motif, geometrical frustration will occur, when all three AFM couplings, or two FM ones together with one AFM, are present. The geometrical frustration is not present in a trimer with two AFM couplings together with one FM coupling or when all three couplings are FM[14,15]. Interest in such systems is not only confined to the geometrical frustration, but also to some fundamental phenomena in magnetism like spin-canting, metamagnetic transition and spin-flop (SF) transition. The net moments of the spins aligned antiparallelly by a weak AF interaction, which is characterized by the metamagnet using a large external field, could overwhelm the weak AFM interaction and turn the system to a ferri-, ferro-, or weak ferromagnetic state[16-19]. Some studies discuss the simultaneous presence, for example, the spin-canting and SF or, spin-canting and metamagnetism in one system[20-25]. Only few reports outline all three phenomena in one system[26,27].

The relevant question is, which underlying magnetic spin model is the appropriate one for such compounds. The standard procedure of developing such a spin model is, to fit bulk magnetic susceptibility data with the respective assumed magnetic model. This procedure may result not only in one unique solution. This is due to the rather insensitive fitting parameter dependence or due to the rather insensitive nature of the magnetic susceptibility on the microscopic details of the magnetic models. Therefore, the microscopic understanding of the details of the magnetic behaviour of the compounds is essential for the comprehension of the mechanism of the exchange interactions. In comparison to W. Guo et al.[13], the structure investigations for compound (1) show that the AFM interaction in this system is quasi-2D. Figure 1a) shows the 2D magnetic network in *bc*-plane. A trimer unit is connected to another one in *a*-direction (see Figure 1b)) and together with their connection in the *bc*-layer, they form a double layer of trimers. Such double layers are separated by Cs atoms, which are not shown for clarity.

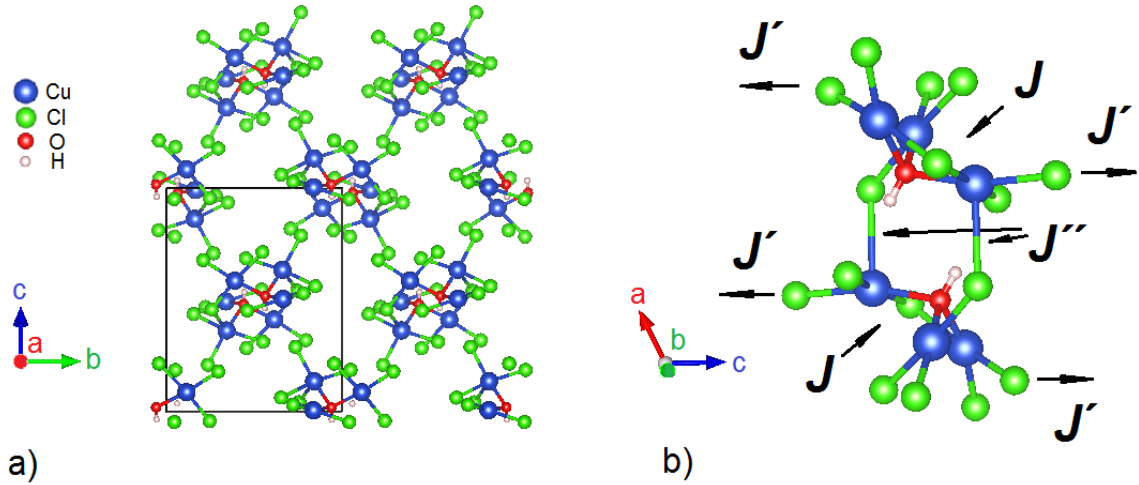

**Figure 1.** a) Crystal structure for $Cs_3Cu_3Cl_8OH$, reduced view for *bc*-plane, b) schematic magnetic exchange interactions between [$Cu^{2+}$] ions within the *bc*-plane

By considering a similar interaction (*J* – intratrimer interaction) between the magnetic ions $Cu^{2+}$ in such a triangular lattice (Figure 1b, $J1 = J2 = J3 = J$)[28-30], it can be assumed that an intertrimer interaction (*J'*) in *bc*-plane with components in *b*- and *c*-direction and also an intertrimer interaction (*J''*) in *a*-direction take place. So far, the experimental results suggest that there is a strong intra-trimer FM coupling and a weak inter-trimer AFM coupling exists [13]. Using neutron diffraction on single crystals, it should be possible to work out the type of the magnetic order at low temperatures of (1), to understand its magnetic structure in more detail.

In this paper, the experimental details are presented in Sec. II. The results of crystal growth, structural characterization, magnetic properties and the first investigation using single crystal neutron diffraction of the spin trimer compounds (1) and (2) are described in Sec. III, followed by a conclusion and outlook.

## II. EXPERIMENTAL SECTION

- **Chemical details for crystal growth of new $Cs_3Cu_3Cl_{8-x}Br_xOH$**

Single crystals of (1) and (2) were grown from aqueous solution using the evaporation method[31]. The crystal growth has been performed at 295 K. For compound (1), the crystalline reagents CsCl (Roth) and $CuCl_2·H_2O$ (Merck) in a molar ratio 3:1 (for growth temperature 281 K) and 5:1 (for growth temperature 295 K) were used. The required solute compositions for the growth of (2) were prepared using the following crystalline reagents: CsCl (Roth), CsBr (VWR), $CuCl_2·2H_2O$ (Merck)[31]. The chemical composition of the samples was determined via energy dispersive x-ray analysis (EDX) using a Zeiss Leo 1530. The differences between the Cl/Br ratios in the solutions and in the grown crystals were determined. These are within the uncertainty range of the EDX analysis (max. ±1at%). Therefore, the data presented here are marked with the corresponding initial compositions of the solutions.

- **X-ray crystallography**

For the single crystal structure determination, a STOE IPDS-II diffractometer with a Genix microfocus tube and with mirror optics was used. All crystals were measured at 173 K. Indexing, integrating and absorption correction were processed in the X-AREA program system[32]. The structures were solved by direct methods and refined with full-matrix least-squares techniques on $F^2$ using the program SHELXL[33].

- **Magnetic susceptibility measurements**

The magnetic susceptibility of these compounds was measured using a MPMS (Quantum Design) in a temperature range from 1.8 K to 300 K and different magnetic fields up to 7 T. The magnetic fields were oriented parallel and/or perpendicular to the *b*-axis. The background signal including the quartz sample holder and GE varnish (used to fix the sample) was measured at the same field conditions. All experimental magnetic data were corrected for diamagnetism of the constituent atoms (Pascal's tables)[34].

- **Neutron diffraction**

Single crystal neutron diffraction experiments were carried out on the thermal-neutron diffractometers ZEBRA(TRICS) and on the cold neutron powder diffractometer DMC at SINQ of Paul Scherrer Institute (PSI) in Villigen, Switzerland[35-37]. The investigation on DMC was performed in single crystal mode. A monochromatic neutron beam with wavelengths 2.317 Å with PG-filter (ZEBRA/TRICS), 2.458 Å with PG-filter and 4.507 Å with Be-filter (both at DMC) were produced through a $PG_{002}$-monochromator. Normal beam geometry was used. The diffracted signal has been collected with a single $^3$He–tube detector and an area detector. The crystals were cooled down to 50 mK.

III. RESULTS AND DISCUSSIONS

**Crystal growth of new $Cs_3Cu_3Cl_{8-x}Br_xOH$**

The mixed system $Cs_3Cu_3Cl_{8-x}Br_xOH$ was formed by substitution of Cl by Br atoms. The maximum of the substitute value of Br is, up to now, x = 1.1 [31]. The crystallization of this mixed system is performed from the quaternary system $CsCl$-$CsBr$-$CuCl_2$-$H_2O$. For crystal growth a temperature of 295 K and 281 K with a constant temperature profile was applied. At 295 K, the two phases of $Cs_3Cu_3Cl_{8-x}Br_xOH$ and $Cs_2CuCl_{4-x}Br_x$ have been observed.

The crystallization of (1) was performed from the quasi-ternary system $CsCl$-$CuCl_2$-$H_2O$. A temperature of 295 K was used. The molar ratio 5:1 of CsCl and $CuCl_2 \cdot 2H_2O$ was selected for this growth temperature, which results in the formation of the two phases $Cs_3Cu_3Cl_8OH$ and $Cs_2CuCl_4$. The crystallization of (1) lasts between four weeks and six months, depending on the evaporation rate. The volume of the crystallized material is small. The colour of the crystals is dark red. The value of the Cs, Cu, Cl elements of the samples was determined by EDX and showed the following results in at.%: Cs-21.36±0.13, Cu-21.04±0.20, Cl-57.60±0.21. EDX's quantitative oxygen and hydrogen analysis is difficult and unprecise. Therefore, the ratio of the elements Cs:Cu:Cl (1:1:2.7), which are characteristic for this compound, are examined.

Regarding the new phase (2), the volume of the crystallized material is also small. At 281 K, three phases of $Cs_3Cu_3Cl_{8-x}Br_xOH$, $Cs_2CuCl_{4-x}Br_x$ and $Cs_3Cu_2Cl_{7-x}Br_x \cdot 2H_2O$ were detected[31]. The volume of the crystallized material of $Cs_3Cu_3Cl_{8-x}Br_xOH$ with x = 0.4, 0.6, 0.8, 1.0, 1.1, grown at 281 K is larger than that of compounds grown at 295 K. The colour of the crystals is also dark red. In the following, only the investigation results of the single crystals with x = 0.4 are outlined. The value of the Cs, Cu, Cl and Br elementsof the samples was determined by EDX and showed the following results in at.%: Cs-21.50±0.14, Cu-21.44±0.23, Cl-53.76±0.20 and Br-3.30±0.07. Here, the characteristic ratio of the elements Cs:Cu:(Cl+Br) (1:1:2.7) remains unchanged. As a result, the chemical composition shows a value for the Br concentration of x = 0.47. Since the uncertainty range of the EDX analysis is max. ±1at%, we use in this paper for the Br concentration the value x=0.4. Single crystals of mixed system $Cs_3Cu_3Cl_{8-x}Br_xOH$ are shown in Figure 2.

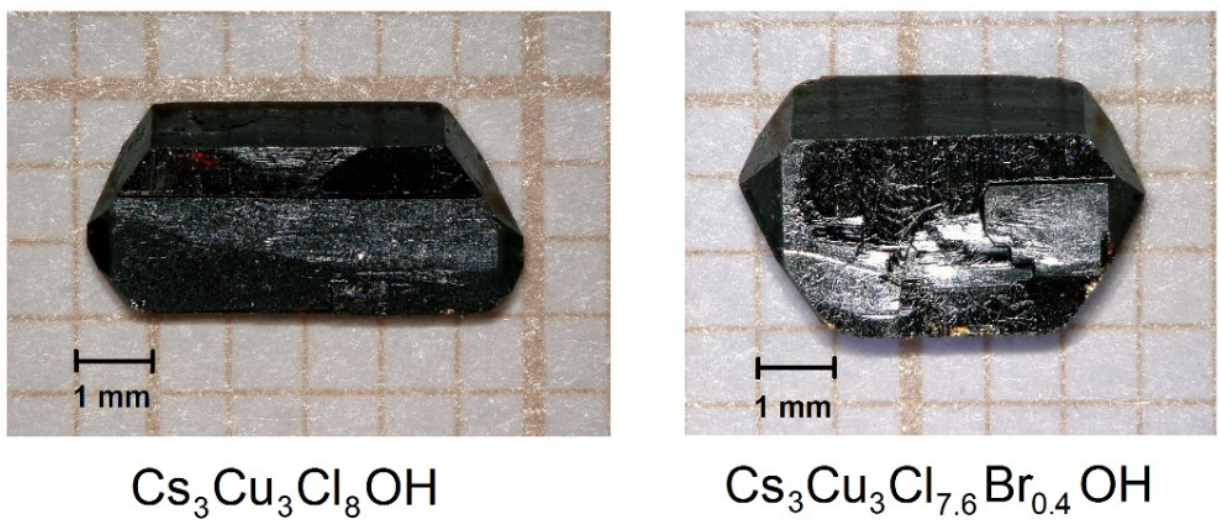

**Figure 2.** Single crystals with Br concentration x = 0 and x = 0.4

**Crystal structure determination**

Both compounds (1) and (2), grown with the same growing method, are monoclinic and crystallize in space group $P2_1/c$. They show small structure differences. In the following, the result of the investigation of the structure features of compounds (1) and (2) are presented, as it is important to understand the relationship between such structure differences and the diverging physical properties of these compounds.

The crystal structure of compound (1) is, in all aspects, similar to the one reported by W. Guo et al., which was grown using the hydrothermal reaction[13]. Both compounds (1) and (2) are isostructural, and, to describe the structural differences, the details of compound (1) and (2) are outlined in the following. A perspective view of the new compound (2) in the *ac* plane is shown in Figure 3a), and a detailed view of the octahedral environment of $Cu^{2+}$ is presented in Figure 3b).

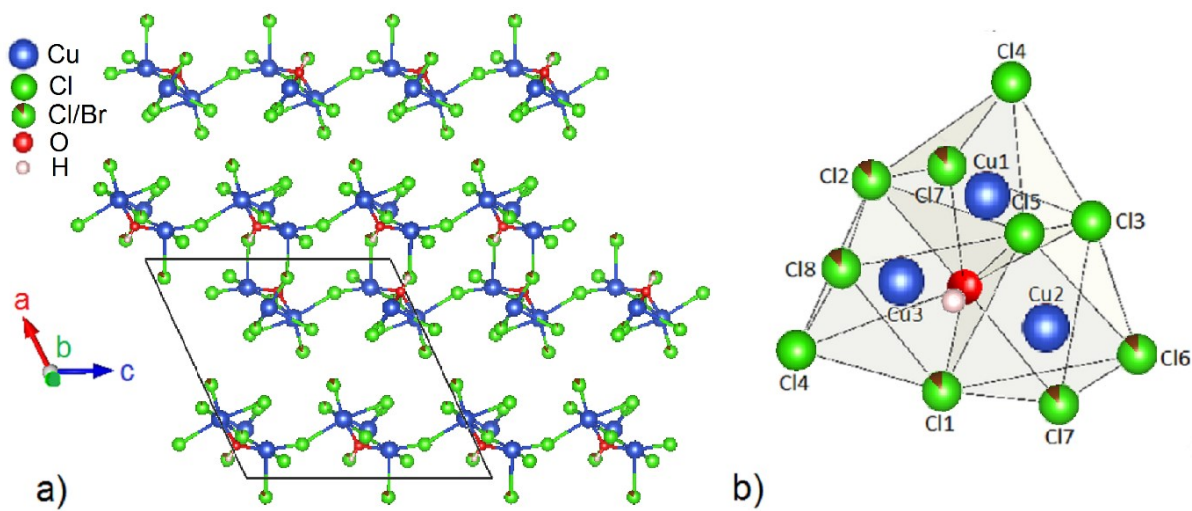

**Figure 3.** a) Perspective view of (2) along the *b*-axis (reduced picture, the Cs atoms are not shown for clarity), b) octahedral environment of $Cu^{2+}$ in (2), in green/brown: crystallographic positions partially occupied by Br

The Cu atoms have an octahedral environment of five Cl/Br and one O atoms and form triangular groups. In the middle of the triangular groups, an O-H bonding can be found. Figure 3b) shows the triangular groups with the $Cu^{2+}$ environment. All octahedrons are distorted, and for each three octahedrons, the shared O-H bonding is shown. This grouping induces an additional distortion, which is overlapped with the Jahn-Teller-distortion. The results of the refinement show that Cl7 is the preferred crystallographic position occupied by Br. In Table 1, the distances from Cu atoms to all vertices of the three octahedrons of (1) and (2) are shown. Therefore, this configuration for (2) leads to different Cu-Cu-Cu distances between 3.074 Å and 3.106 Å in the trimer.

**Table 1.** Distance between Cu atoms to the ligands/vertices of the octahedrons of (1) and (2)

| Atoms | Distances [Å] (1) | Distances [Å] (2) | Atoms | Distances [Å] (1) | Distances [Å] (2) | Atoms | Distances [Å] (1) | Distances [Å] (2) |
|---|---|---|---|---|---|---|---|---|
| Cu(1)–O(1) | 1.963(3) | 1.968(3) | Cu(2)–O(1) | 1.973(3) | 1.978(3) | Cu(3)–O(1) | 1.979(3) | 1.990(3) |
| Cu(1)–Cl(2)/Br(2) | 2.3773(11) | 2.3950(11) | Cu(2)–Cl(5) | 2.3877(11) | 2.3975(11) | Cu(3)–Cl(1)/Br(1) | 2.3098(11) | 2.3220(11) |
| Cu(1)–Cl(3) | 2.3261(11) | 2.3332(11) | Cu(2)–Cl(6)/Br(6) | 2.2288(11) | 2.2504(11) | Cu(3)–Cl(2)/Br(2) | 2.3896(11) | 2.4031(11) |
| Cu(1)–Cl(4) | 2.2356(11) | 2.2463(11) | Cu(2)–Cl(7)/Br(7) | 2.3101(11) | 2.3434(10) | Cu(3)–Cl(8)/Br(8) | 2.2275(10) | 2.2589(10) |
| Cu(1)–Cl(7)Br(7) | 2.7411(11) | 2.7675(10) | Cu(2)–Cl(1)/Br(1) | 2.7631(11) | 2.7814(11) | Cu(3)–Cl(4) | 2.7723(11) | 2.7852(12) |
| Cu(1)–Cl(5) | 2.8561(10) | 2.8653(10) | Cu(2)–Cl(3) | 2.8404(10) | 2.8535(11) | Cu(3)–Cl(5) | 2.8352(11) | 2.8475(11) |

For both compounds (1) and (2), each copper atom is coordinated with one oxygen atom with Cu-O distances in the range of 1.96 Å to 1.99 Å. Furthermore, there are three shorter Cu-Cl bondings between 2.2 Å and 2.4 Å and two larger ones with distances between 2.7 Å and 2.9 Å. Compound (1) shows a similar octahedral $Cu^{2+}$ environment. The distances in the trimer of (1) are between 3.064 Å and 3.100 Å and are also very similar to those of compound (2).

The crystal data, the data collection, and the structure refinement details are summarized in Table 2. The results show that only the lattice constants of compound (2) are slightly larger in comparison to (1).

**Table 2.** Selected crystallographic data for compounds (1) and (2)

| Formula after refinement of crystallographic positions | $Cs_3Cu_3Cl_8OH$ | $Cs_3Cu_3Cl_{7.64}Br_{0.36}OH$ |
|---|---|---|
| Formula weight | 889.96 | 906.06 |
| Radiation, $\lambda$ [Å] | MoK$\alpha$, 0.71073 | MoK$\alpha$, 0.71073 |
| Crystal system | monoclinic | monoclinic |
| Space group | $P2_1/c$ | $P2_1/c$ |
| $a$ [Å] | 13.3740(7) | 13.4523(6) |
| $b$ [Å] | 9.6991(4) | 9.7360(4) |
| $c$ [Å] | 13.6209(8) | 13.6929(7) |
| $\alpha$ [°] | 90 | 90 |
| $\beta$ [°] | 114.850(4) | 114.854(4) |
| $\gamma$ [°] | 90 | 90 |
| Volume [Å$^3$] | 1603.25(14) | 1627.28(14) |
| Z | 4 | 4 |
| $D_{calcd}$ [Mg cm$^{-3}$] | 3.687 | 3.698 |
| F(000) | 1588 | 1614 |
| Absorption coefficient | 11.960 | 12.613 |
| Crystal size [mm] | 0.09×0.09×0.03 | 0.19x0.13x0.08 |
| Total reflections ($R_{int}$) | 39392(0.0922) | 26851(0.0617) |
| Data/restraints/parameters | 4497/0/136 | 4541/0/142 |
| GOF on $F^2$ | 1.009 | 1.112 |
| $R_1$, $wR_2$ [$I > 2\sigma(I)$] | 0.0300, 0.0644 | 0.0442, 0.1160 |
| $R_1$, $wR_2$ (all data) | 0.0401, 0.0674 | 0.0454, 0.1175 |
| Largest diff peak and hole [e Å$^{-3}$] | 1.095, -1.983 | 4.124, -1.883 |

The cif files containing all crystallographic data have been deposited with the CCDC. The reference numbers are 1950448 for (1) and 1950447 for (2).

**Magnetic properties**

Magnetic susceptibility indicates that the magnetic properties of (1) are dominated by the interaction among magnetic Cu ions including weak intertrimer and intratrimer interactions. As shown in Figure 4a), the Neel temperature $T_N$ is at $T_{N1} = (2.23 \pm 0.5)$ K. Towards lower $T$, the drop of the susceptibility shows an AFM behaviour. With respect to the trimer model, the ground state is equal to $S = \frac{1}{2}$, if the magnetic susceptibility decreases with the temperature dropping to $T = 0$ K. For compound (2) $T_{N2} = (2.70 \pm 0.5)$ K (see Figure 4c)). No noticeable difference is found between zero-field-cooled (ZFC) and field-cooled (FC) susceptibilities. The Curie-Weiss temperature $\theta_1$ can be extracted from the inverse susceptibility $\chi^{-1}_{mol}(T)$ between 50 K and 160 K for (1) (see Figure 4a)), which is similar to the value published by W. Guo et al 13. The calculation resulted in $C_1 = 17.950(1) \cdot Km^3 \cdot mol^{-1}$ and $\theta_1 = (6.76 \pm 0.35)$ K. It is assumed that the small kink between 160 K and 170 K was caused by changing the steps during the measurement. The Curie-Weiss temperature $\theta_2$ for compound (2) can be extracted from $\chi^{-1}_{mol}(T)$ (see Figure 4c)) between 200 K and 300 K. The respective calculation resulted in $C_2 = 13.650(1)$ $Km^3 \cdot mol^{-1}$ and $\theta_2 = (14.87 \pm 0.35)$ K. This means that a Br doping leads to a change to a dominant value of the FM interaction of (2). The parameter $f = |\theta_{cw}|/T_N$ has a value for (1) of 3.03 and for (2) of 5.51 and corresponds to frustration[4].

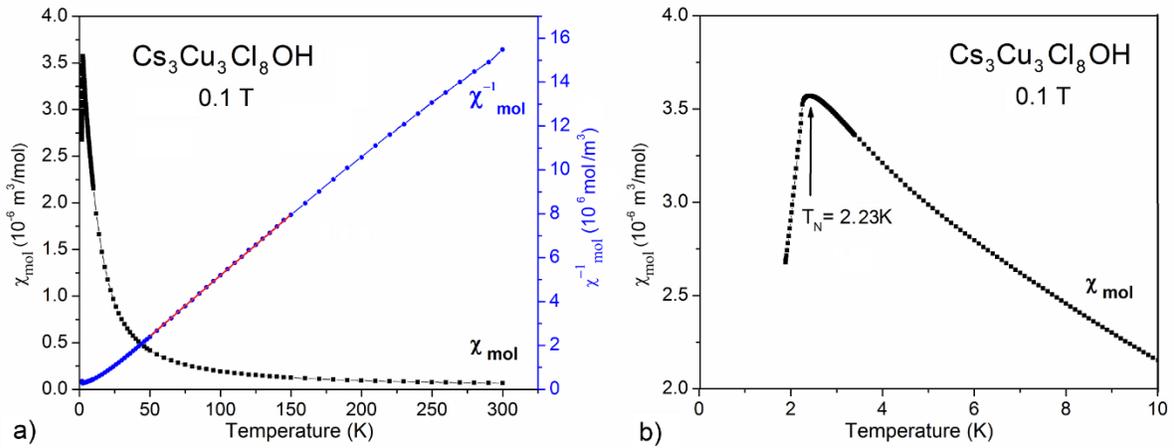

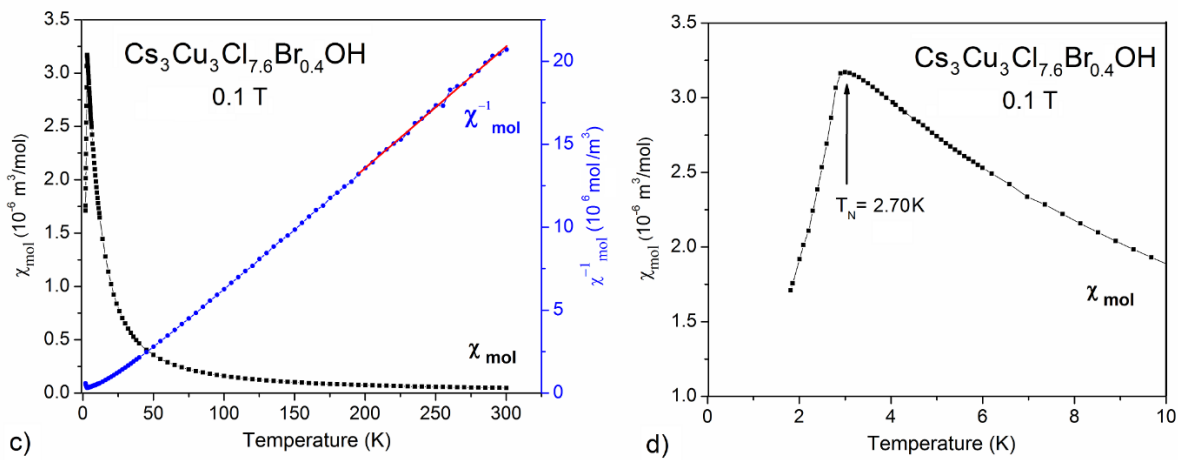

**Figure 4.** Magnetic susceptibility $\chi_{mol}(T)$ vs. $T$ and the inverse susceptibility $\chi^{-1}_{mol}(T)$ depending on the temperature in the magnetic field 0.1 T: a) of compound (1), b) detailed view of the susceptibility maximum for (1), c) of compound (2), and d) detailed view of the susceptibility maximum for (2); the linear fit of the Curie-Weiss behaviour is marked red for both compounds

As presented in Figures 4b) and d), both, the maximum of the susceptibility being shifted to a higher temperature with an increase of the Br concentration from 2.23 K to 2.70 K and the reduction of the value of $\chi_{mol}(T_{max})$ indicate that the magnetic coupling is modified. The value of $(\chi_{mol}(T)*T)^{1/2}$ increases depending on the temperature and realizes a maximum at 18.1 K for (1) and at 17.67 K for (2) (see Figure A in ESI). These maxima show that an intra-trimer FM coupling exists. With lowering the temperature after having reached the respective maximum, the value of $(\chi_{mol}(T)*T)^{1/2}$ decreases rapidly. A similar behaviour was reported for complexes with FM interactions[15].

The effective magnetic moment of the $Cu^{2+}$ ions was determined at 1.95(4) $\mu_B$ parallel to the b-axis for compound (1) and at 1.81(1) $\mu_B$ for compound (2), which are slightly larger than the theoretical value of 1.732 $\mu_B$ for $Cu^{2+}$ ions, obtained from $\mu_{eff}^2 = g^2S(S+1)$ with respect to $S = \frac{1}{2}$ and $g = 2$. From the effective magnetic moment along the b-axis of (1), the calculated $g_{com1} = 2.26$ and for (2) $g_{com2} = 2.09$. Both effective magnetic moments confirm the orbital moment contribution of the $Cu^{2+}$ ions in such distorted octahedrons[38-40].

For both compounds (1) and 2), the magnetization in a magnetic field shows an anisotropy in different crystallographic directions.

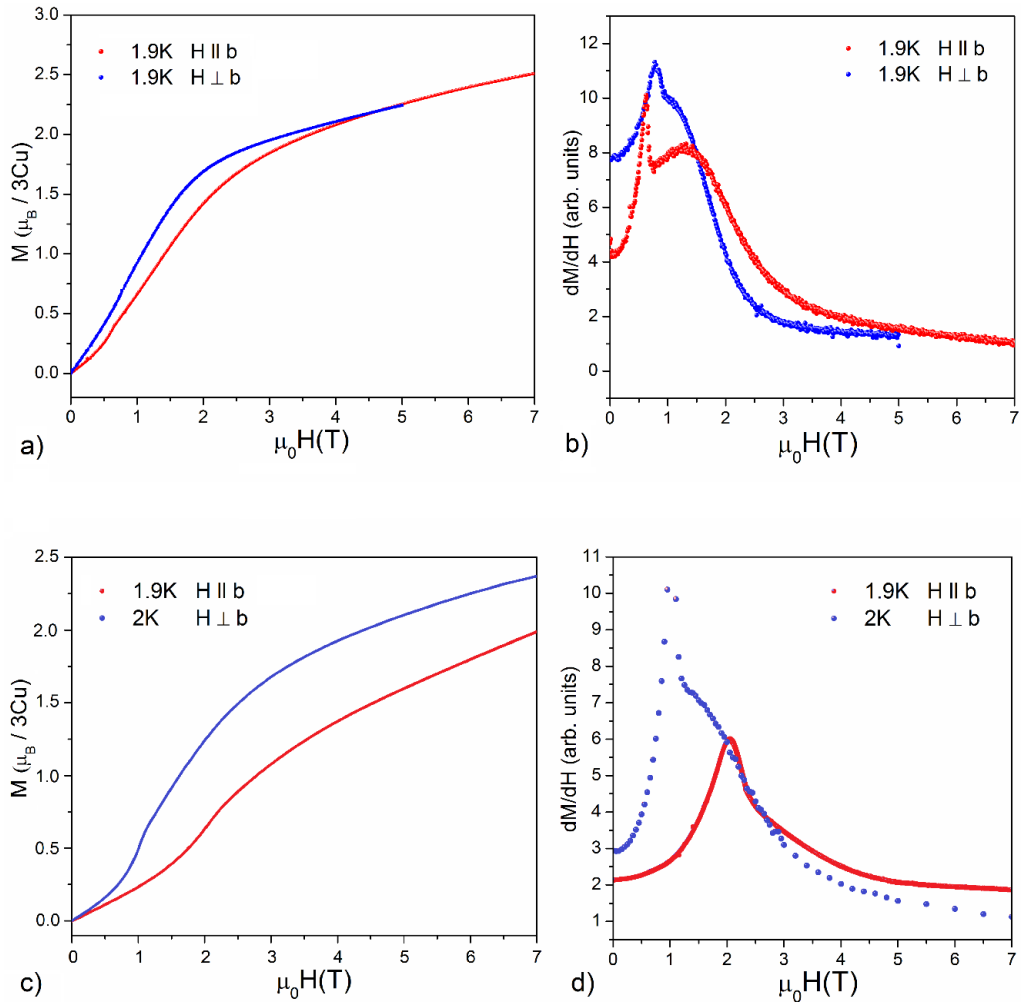

**Figure 5.** Magnetization with a magnetic field parallel and perpendicular to the *b*-direction: a) of (1), b) dM/dH vs. H at 1.9 K, c) of (2) at 1.9 K and 2.0 K, and d) dM/dH vs. H also at 1.9 K and 2.0 K below the temperature of the AFM phase transition

Figures 5a) and c) present the results of the magnetization measurements, perpendicular and parallel to the b-axis. Both compounds do not reach their saturation value at a temperature of 1.9 K and a magnetic field of 7 T.

This can be seen in the derivation curve dM/dH vs. H in Figures 6b) and d), whose values at 7 T are not zero. At 0.61 T for (1) and 2 T for (2) for *b* parallel H and at 0.77 T for (1) and 1 T for (2) for *b* perpendicular H, the slope change of dM/dH vs. H is due to a SF transition in another AFM phase. The low value of the magnetic field for the SF transition reflects the small value of intertrimer interaction (*J'*) in *c*-direction or intratrimer interactions (*J''*) in *a*-direction for (1), whereas the exact direction could not yet be determined. For (2) in comparison to (1), the value for SF transition is slightly larger and originates from a change of Cl/Br atoms in the $Cu^{2+}$ environment. The derivation curve shows clearly that each maximum of (1) and (2) has a shoulder, which confirms the complex spin structure of the investigated compounds[27]. For compound (2), the behaviour of dM/dH vs. H differs from that of compound (1) at a temperature of 1.9 K. In general, the magnetic anisotropy plays an important role in the SF transition of the investigated compositions (1) and (2) [26].

    The susceptibility data depending on temperature and magnetic field are presented in Figure 6a). This shows that the maximum of the susceptibility, which was observed at around 2.23 K in zero field, can be suppressed in an applied field of 1.1 T. The details are shown in Figure 6b). The same effect was observed for compound (2) also in an applied field of 1.2 T (see Figures 6c) and d)).

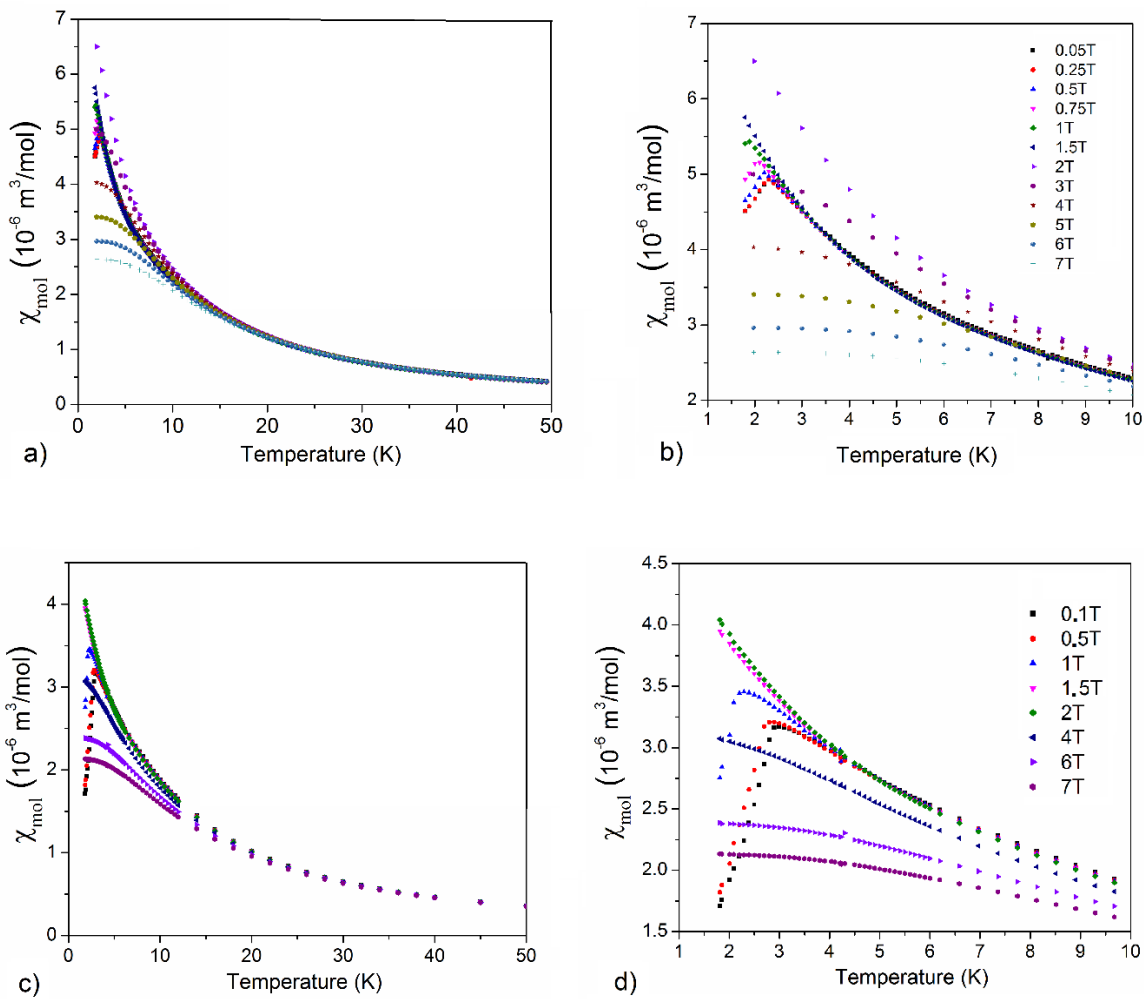

**Figure 6.** Susceptibility depending on temperature and magnetic field parallel to the *b*-direction a) of compound (1), b) detailed view of the susceptibility maximum of compound (1), c) of compound (2), d) detailed view of the susceptibility maximum of compound (2)

As shown in Figure 6, the susceptibility behaviour of each compound changes above the respective critical field. Such susceptibility behaviour is characteristic for weak coupled spin systems[41].

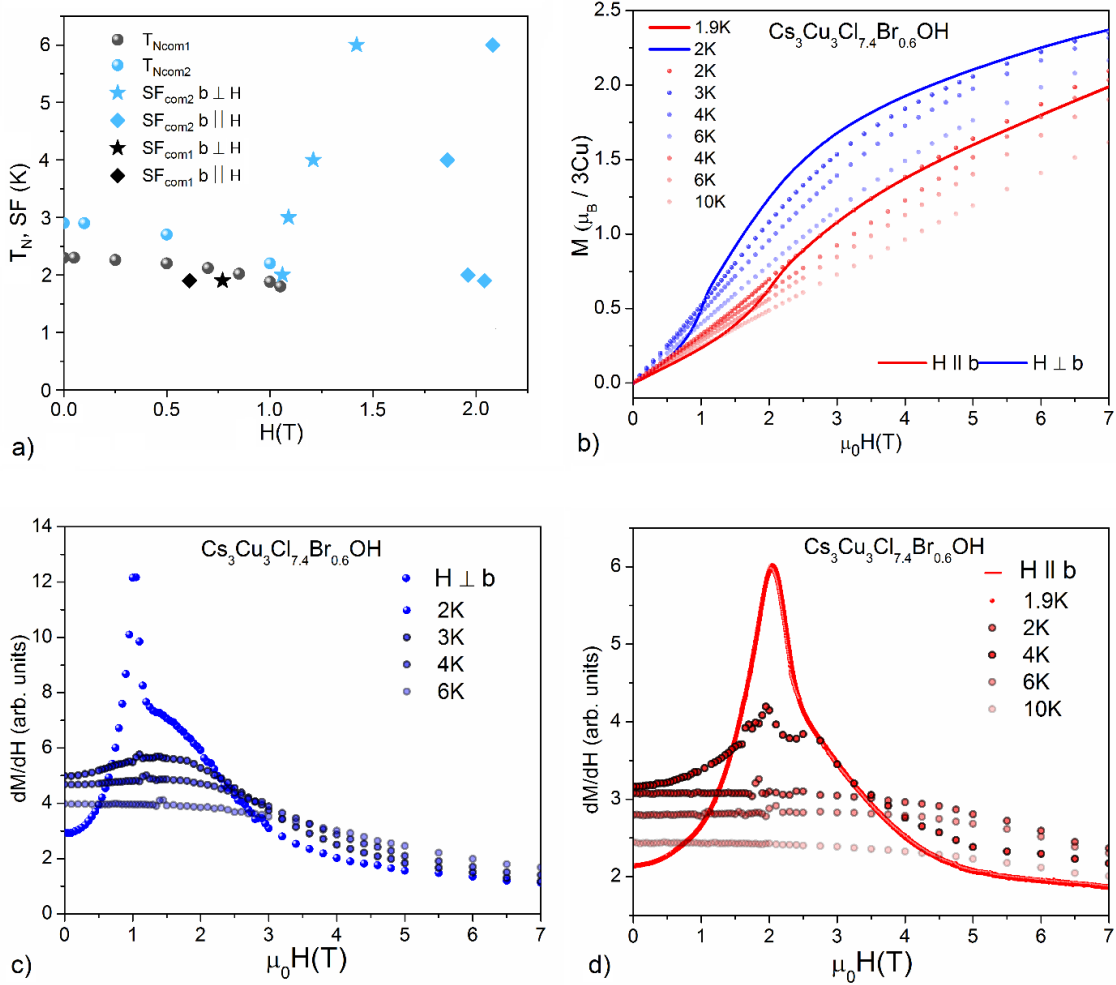

**Figure 7.** a) Phase diagram of the magnetic order depending on temperature and magnetic field for (1) as black circles and for (2) as blue circles; the black star and the black diamond mark the SF transition for (1) and the blue stars and the blue diamonds such of (2), b) Magnetisation with a magnetic field parallel and perpendicular to the b-direction measured between 1.9 K – 10 K of (2), c) dM/dH vs. H perpendicular to *b* of (2), d) dM/dH vs. H parallel to *b* of (2)

Figure 7a) shows the results of the susceptibility data for both compounds, which are evidenced for an AFM phase transition, respectively at $(2.23 \pm 0.5)$ K for (1) and at $(2.70 \pm 0.5)$ K for (2), and in zero field for both compounds. These transitions can be suppressed in a magnetic field at 1.1 T for compound (1) and at 1.2 T for compound (2). The results of magnetisation with a magnetic field at temperatures between 1.9 K and 10 K are presented in Figure 7b). The derivation curves dM/dH vs. H of (2) (see Figure 7c) and d)) show that the SF transition is seen up to 6 K. In summary, for compound (1), the SF transition happens in the AFM phase below the temperature of the long-range magnetic order using a magnet field at 0.61 T and 0.77 T depending on the measured direction to the magnetic field. The SF transition of (2) occurs after suppressing the AFM transition using a magnet field at 2 T (1.9 K) for the measured *b*-direction parallel to the magnetic field.

Neutron powder diffraction experiments were performed on (1). Figure 8a) shows diffraction patterns at 1.5 K and 5 K, as well as their difference pattern (blue curve). The difference curve does not show any features and, therefore, no structural or magnetic phase transition could be detected. Each 2θ point was measured 6 hours. Thereafter, a single crystal neutron diffraction experiment on (1) was executed in the temperature region between 2.50 K and 1.75 K. Figure 8b) shows the temperature evolution of the integrated intensity of the (0 0 2) reflection for (1), which indicates that the magnetic ordering takes place at $T_{N1}$ = 2.23(3) K. The magnetic signal appears on the top of the nuclear reflections. Each point was measured 20 minutes. With regard to the investigated temperature region, only the neutron single crystal diffraction shows a small, significant rise of the integrated intensity curve, when lowering the temperature. The aforementioned investigations show that a lower investigation temperature is needed to receive more information about the magnetic behaviour of compound (1).

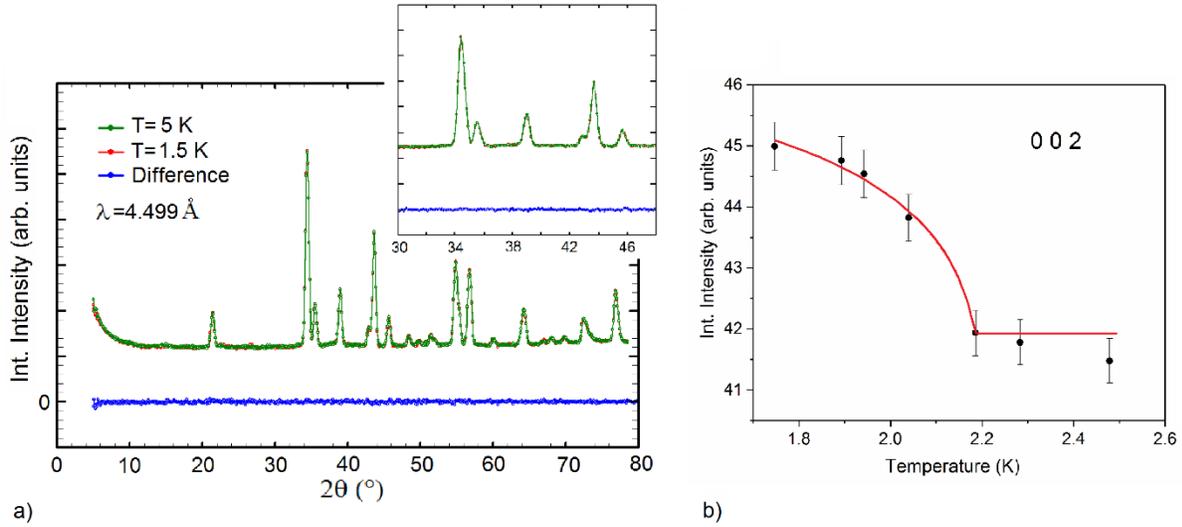

**Figure 8.** a) Difference of the intensity in zero field between 1.5 K and 5.0 K of composition (1) from the neutron powder diffraction, measured on DMC, and b) *T*-dependence of the intensity for (0 0 2) of compound (1), measured as neutron single crystal diffraction on Zebra between 2.5 K and 1.75 K, fit of power-law behaviour marked in red

To increase the probability to detect AFM reflections, neutron single crystal diffraction experiments on DMC, mapping the *ab*-plane at different temperatures, have been executed. This method allows for a measurement of reciprocal planes but is not suitable for extracting reliable integrated peak intensities. A counting at 3 K above the temperature of the phase transition (at around 2.23 K) and at 100 mK was implemented for 24 hours for each temperature point for 2θ from 10° to 90°. Figure 9a) shows the intensity difference in the *ab*-plane at 100 mK and at 3 K, respectively. No additional reflections have been observed on half integer positions between the nuclear ones down to 100 mK. The mapping of the intensity difference in the *ab*-plane (Figure 9a)) shows only qualitative results. Despite AFM ordering the magnetic signal is only visible at the positions of nuclear reflections, which points to an AFM arrangement of the magnetic moments within the crystallographic unit cell. The background value is around 5 counts per second, and this shows a very small background for this measurement.

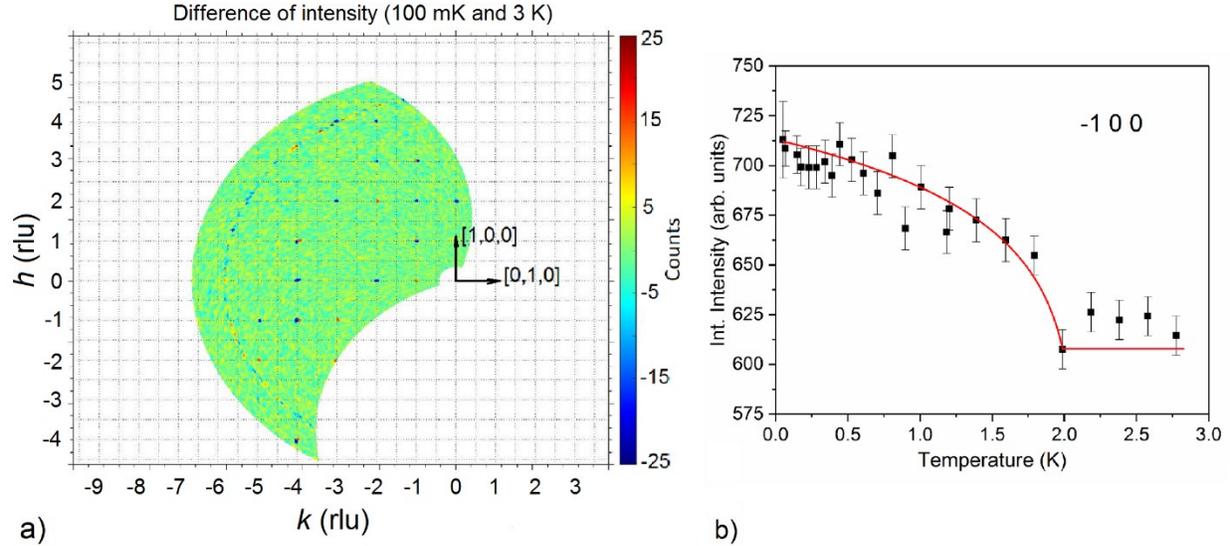

**Figure 9.** a) Difference of the intensity in zero field between 100 mK and 3 K of composition (1) as a map of the (*h k 0*) plane; the step (Δϕ) in these measurements was 0.2°; measured on DMC, and b) *T*-dependence of the integrated intensity for (-1 0 0), measured on Zebra, fit of power-law behaviour marked in red; both executed as neutron single crystal diffraction experiments

Figure 9b) shows the temperature evolution of the integrated intensity of the (-1 0 0) reflection for (1) in the temperature range 50 mK to 3 K, which indicates the onset of magnetic ordering at $T_{N1}$ = 2.12(3) K. The solid line represents the fit of the power-law behaviour $I \propto (T_N - T)^\beta$ to the integrated magnetic peak intensity of (1). The critical exponent $\beta$ is associated with the order parameter of the system. In this case, no values of $\beta$ are compared since no theoretical comparison exists for such systems.

The structures of (1) and (2) show the symmetry $P2_1/c$. They include 12 Cu atoms in their unit cells. Three crystallographically independent Cu atoms form trimers, which are present in the unit cell. Several magnetic space groups, based on the $P2_1/c$ nuclear space group, exist that are compatible with AFM. The structure is without requesting a supercell. This magnetic order is then exploiting the magnetic contributions to the nuclear reflections, as they have been observed below $T_N$ (see Figure 8a) and 9a)). The magnetic structure cannot yet be finally solved, as only two magnetic reflexes (-1 0 0) and (0 0 2) have been investigated so far. The results show that the magnetic moment consists of components in all three directions (x, y, z). Based on the results of the experiments, the possible magnetic space group could be one of the following: $P2_1/m$, $P2_1'/m'$, $P2_1'/m$ or $P2_1/m'$.

To understand the nature of the magnetic behaviour of compounds (1) and (2), it would be beneficial, to prepare a spin model for the fitting of the investigated magnetic data. Such a model should also answer the question, whether the Hamiltonian equation for the equidistant trimer with respect to the weak interaction between the trimers is sufficient to describe the interactions in (1) and (2). In compound (1), the corresponding distances between Cu–Cu are 3.064 Å, 3.088 Å, and 3.099 Å (see Figure 10a)). For answering this question, it is important to perform a theoretical calculation for equidistant and nonequidistant trimers. Figure 10b) shows that each Cu trimer is connected to the next one through the Cl atom in *c*-direction: Cu1-Cl4=2.2357(12) Å and Cu3-Cl4=2.7722(11) Å, and through two Cl atoms in *a*-direction: Cu1-Cl7=2.7412(11) Å and Cu2-Cl7=2.3101(10) Å. In particular, this indicates that the exchange coupling $J'$ along the Cu chains arises from the Cu1-Cl4-Cu3 hybridization, but not exactly in *c*-direction (in *bc*-plane). The exchange coupling $J''$ perpendicular to the Cu chains arises from Cu1-Cl7-Cu2, which is close to the *a*-direction.

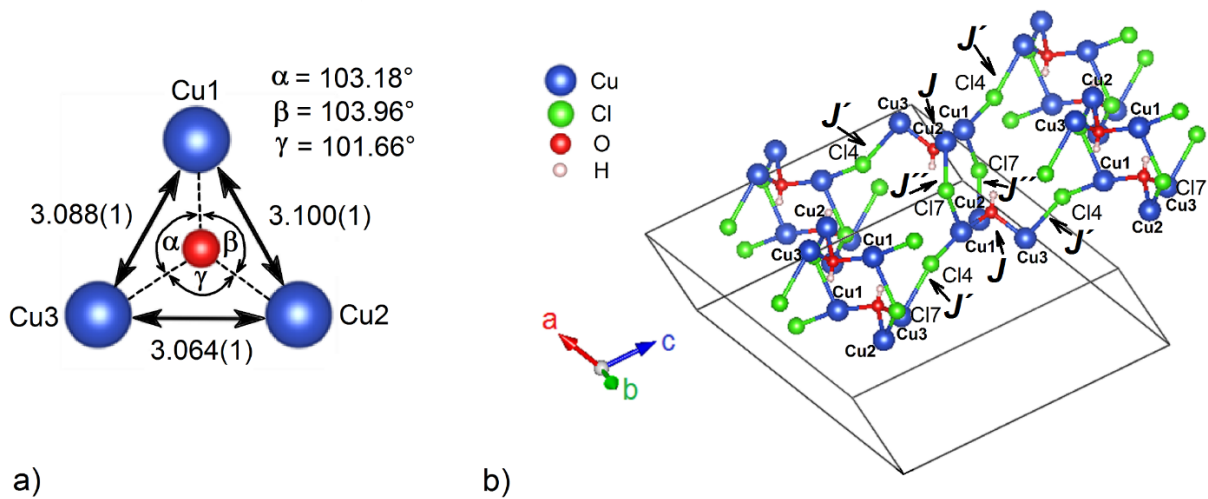

**Figure 10.** a) Trimer unit, Cu-Cu distances in Å and b) distances in *a*- and *c*-direction, which are relevant for magnetic interaction *J, J′* and *J″* in compound (1), for compound (2) the crystallographic position Cl7 is partially occupied by Br

The aforementioned construction shows a structure of a trimer-layer unit (see Figure 10b)). These units are separated through Cs atoms from each other with a distance between the trimer-layer units of around 7 Å, which is too long for any magnetic interaction. In case of a small Br concentration, the Cl- ions are replaced by Br- ions in the same crystallographic positions of the Cu octahedrons (see Figure 3)). In compound (2), the corresponding distances between Cu–Cu are 3.074 Å, 3.100 Å, and 3.106 Å and the corresponding distances between Cu-Cu are very similar to compound (1). Each trimer is also connected to the next one through the Cl atom in *c*-direction: Cu1-Cl4=2.2464(1) Å and Cu3-Cl4=2.7851(2) Å. The crystallographic position Cl4 is not occupied by Br and the distances are only slightly larger as for compound (1). In *a*-direction, the distances of Cu1-Cl7/Br7=2.7675(1) Å and Cu2-Cl7/Br7=2.3434(2) Å are larger, because the crystallographic position Cl7 is partially occupied by Br. This results in modifying the exchange coupling in this direction for compound (2).

## IV. CONCLUSION AND OUTLOOK

The new mixed system $Cs_3Cu_3Cl_{8-x}Br_xOH$ with weakly coupled triangles shows that the single crystals can be grown with Br concentration from $x = 0.4$ to $1.1$. The two investigated compounds (1) and (2) are isostructural. They crystallize in the monoclinic space group $P2_1/c$. For the investigated mixed composition, the Cl- ions are replaced by Br- ions in the same crystallographic positions of the Cu octahedrons. The differences of distances and angles are very small, but, nevertheless, those are important to detect distinct differences in the magnetic properties of both compounds. The temperature of the AFM phase transition increases from $T_{N1} = (2.23 \pm 0.5)$ K to $T_{N2} = (2.70 \pm 0.5)$ K with a Br concentration of $x = 0.4$. But the phase transition can be suppressed in magnetic field for both compounds, for (1) at 1.1 T and for (2) at 1.2 T. Measurements of the magnetic susceptibility and magnetization also in a- and c-directions have been beneficial to get more information for understanding the magnetic exchange interactions in (1) and (2). In addition, the crystal growth and investigation of compounds with a higher Br concentration are essential, to determine the changes in the magnetic behaviour of such compounds under the condition that the structure remains isostructural.

Neutron single crystal diffraction shows that more intensity can be seen on the top of the nuclear reflections below the $T_{N1} = 2.12(3)$ K phase transition temperature. This is confirmed with bulk measurements, which show AFM phase transition at low temperature. The next step will be the investigation in high magnetic field. The saturation field $B_S$ of (1) and (2) is expected between 15 T and 30 T, which will then have to be compared to theoretical simulations. A density functional theory (DFT) calculation will give interesting additional information about the exchange coupling in the trimer and between the trimers, which is envisaged for the compounds of this mixed system.

**Supporting Information**.

Figure A: The value of $(X_{mol}(T)*T)_{1/2}$ depending on the temperature

**Corresponding Author**

\* Dr. Natalija van Well, Department of Earth and Environmental Sciences, Crystallography Section, Ludwig-Maximilians-University Munich, D-80333 Munich, Germany,

E-Mail: natalija.vanwell@lrz.uni-muenchen.de

**Author Contributions**

Natalija van Well: Conceptualization, Formal analysis, Funding acquisition, Investigation, Methodology, Project administration, Validation, Visualization, Writing - original draft, Writing - review and editing

Michael Bolte: Formal analysis, Investigation, Writing - review and editing

Claudio Eisele: Formal analysis, Investigation, Writing - review and editing

Lukas Keller: Formal analysis, Investigation, Resources, Writing - review and editing

Jürg Schefer: Formal analysis, Investigation, Resources, Writing - review and editing

Sander van Smaalen: Methodology, Resources, Validation, Writing - review and editing.

**Funding Sources**

This work was supported by Paul Scherrer Institute, University of Bayreuth, and the Deutsche Forschungsgemeinschaft through the research fellowship for the project WE-5803/1-1 and WE-5803/2-1, SNF under grant number 206021_139082, University of Bayreuth through fellowship A 4576 – 1/3 and the European Community's Seventh Framework Programme under grant agreement n. 290605 (PSI-FELLOW/COFUND).

ACKNOWLEDGMENT

The authors thank Ch. Rüegg from the Department for Research with Neutrons and Muons (PSI), Villigen, S. Ramakrishnan from Department of Condensed Matter Physics and Material Sciences, Tata Institute of Fundamental Research, Mumbai for fruitful discussions. The authors thank also T. Shang and M. Medarde from the Laboratory for Multiscale Materials Experiments (PSI), Villigen for their support during the experiment at the MPMS and the analysis of the data and E. Canevet from Laboratory for Neutron Scattering and Imaging (PSI), Villigen for his support during the experiment at the DMC and the analysis of the data, and M. Heider from Bayerisches Polymerinstitut, Bayreuth for her support. The neutron diffraction experiments were performed on ZEBRA(TRICS) and DMC and orientation of the crystals on ORION at SINQ of PSI, Villigen, Switzerland.

REFERENCES

[1] S. Sachdev, B. Keimer, Quantum Criticality, *Physics Today*, 2011, 64(2), 29-38
[2] M. Kohno, O. A. Starykh, L. Balents, Spinons and triplons in spatially anisotropic frustrated antiferromagnets, *Nature Physics*, 2007, 3, 790 - 795
[3] L. Balents, Spin liquids in frustrated magnets, *Nature*, 2010, 464, 199-208
[4] M. Vojta, Frustration and quantum criticality, *Rep. Prog. Phys.*, 2018, 81, 064501
[5] J. C. Bonner, S. A. Friedberg, H. Kobayashi, D. L. Meier, and H. W. J. Blote, Alternating linear-chain antiferromagnetism in copper nitrate $Cu(NO_3)_2 \cdot 2.5\ H_2O$, *Phys. Rev. B*, 1983, 27, 248
[6] Ch. Rüegg, N. Cavadini, A. Furrer, H.-U. Güdel, K. Krämer, H. Mutka, A. Wildes, K. Habicht and P. Vorderwisch, Bose-Einstein condensation of the triplet states in the magnetic insulator $TlCuCl_3$, *Nature*, 2003, 423, 62–65
[7] Y. Qiu, C. Broholm, S. Ishiwata, M. Azuma, and M. Takano, R. Bewley, W. J. L. Buyers, Spin-trimer antiferromagnetism in $La_4Cu_3MoO_{12}$, *Phys. Rev. B*, 2005, 71, 214439,
[8] M. Azuma, T. Odaka, M. Takano, D. A. Vander Griend, K. R. Poeppelmeier, Y. Narumi, K. Kindo, Y. Mizuno, and S. Maekawa, Antiferromagnetic ordering of S=1/2 triangles in $La_4Cu_3MoO_{12}$, *Phys. Rev. B*, 2000, 62, R3588

[9] M. Azuma, Z. Hiroi, M. Takano, K. Ishida and Y. Kitaoka, Observation of a Spin Gap in SrCu$_2$O$_3$ Comprising Spin-1/2 Quasi-1D Two-Leg Ladders, *Phys. Rev. Lett.* 73, 3463 (1994)

[10] M. Drillon, M. Belaiche, P. Legoll, J. Aride, A. Boukhari, and A. Moqine, 1D ferrimagnetism in copper(II) trimetric chains: Specific heat and magnetic behavior of A$_3$Cu$_3$(PO$_4$)$_4$ with A = Ca, Sr, *J. Magn. Magn. Mater.*, 1993, 128, 83

[11] A. A. Belik, A. Matsuo, M. Azuma, K. Kindo, and M. Takano, Long-range magnetic ordering of S = 1/2 linear trimers in A$_3$Cu$_3$(PO$_4$)$_4$ (A = Ca, Sr, and Pb), *J. Solid State Chem.*, 2005, 178, 709

[12] M. Ghosh, K. Ghoshray, B. Pahari, R. Sarkar, and A. Ghoshray, 31P NMR of trimer cluster compound Sr$_3$Cu$_3$(PO$_4$)$_4$, *J. Phys. Chem. Solids*, 2007, 68, 2183

[13] W. Guo, Y. Tang, S. Zhang, H. Xiang, M. Yanga, Z. He, Synthesis, structure and magnetic properties of hydroxychlorides A$_3$Cu$_3$(OH)Cl$_8$ (A = Cs, Rb) with isolated tricopper *CrystEngComm,* 2015, **17**, 8471-8476

[14] J. E Greedan, Geometrically frustrated magnetic materials, *J. Mater. Chem.*, 2001, 11, 37-53

[15] B. Sarkar, M. S. Ray, Y.-Z. Li, Y. Song, A. Figuerola, E. Ruiz, J. Cirera, J. Cano, A. Ghosh, Ferromagnetic Coupling in Trinuclear, Partial Cubane Cu(II) Complexes with a µ3-OH Core: Magnetostructural Correlations, *Chem. Eur. J.*, 2007, 13, 9297 – 9309

[16] R. L. Carlin, *Magnetochemistry*; Springer-Verlag: Berlin, 1986

[17] M. Kurmoo, H. Kumagai, M. A. Green, B. W. Lovett, S. J. Blundell, A. Ardavan, J. Singleton, Two modifications of layered cobaltous terephthalate: Crystal structures and magnetic properties, *J. Solid State Chem.*, 2001, 159, 343-352

[18] E. Q. Gao, Z. M. Wang, C. H. Yan, From manganese(II)-azido layers to a novel three-dimensional molecular magnet: spin canting and metamagnetism, *Chem. Commun.*, 2003, 14, 1748-1749

[19] H. Z. Kou, S. Gao, B. W. Sun, J. Zhang, Metamagnetism of the First Cyano-Bridged Two-Dimensional Brick-Wall-like 4f-3d Array, *Chem. Mater.,* 2001, 13, 1431-1433

[20] J. A. Zora, K. R. Seddon, P. B. Hitchcock, C. B. Lowe, D. P. Shum, R. L. Carlin, Magnetochemistry of the Tetrahaloferrate(III) Ions. 1. Crystal Structure and Magnetic Ordering in Bis[4-chloropyridinium tetrachloroferrate(III)]-4-Chloropyridinium Chloride and Bis[4-bromopyridinium tetrachloroferrate(III)]-4-Bromopyridinium Chloride, *Inorg, Chem.*, 1990*, 29,* 3302-3308

[21] J. A. Schlueter, J. L. Manson, K. A. Hyzer, U. Geiser, Spin Canting in the 3D Anionic Dicyanamide Structure (SPh$_3$)Mn(dca)$_3$ (Ph = Phenyl, dca = Dicyanamide), *Inorg. Chem*., 2004, 43, 4100-4102

[22] M. H. Zeng, W. X. Zhang, X. Z. Sun, X.M. Chen, Spin Canting and Metamagnetism in a 3D Homometallic Molecular Material Constructed by Interpenetration of Two Kinds of Cobalt(ii)-Coordination-Polymer Sheets, *Angew. Chem., Int. Ed.*, 2005, 44 (20), 3079-3082

[23] A. Herweijer, W. J. M. de Jonge, A. C. Botterman, A. L. M. Bongaarts, J. A. Cowen, Magnetic Studies of the Canted Ising Linear Chain $CsCoCl_3·2H_2O$, *Phys. Rev. B*, 1972, 5, 4618-4630

[24] K. Kopinga, Q. A. G. van Vlimmeren, A. L. M. Bongaarts, W. J.M. de Jonge, Some Magnetic Properties of the Pseudo one-Dimensional Ising-Like Magnetic Systems $CsCoCl_3·2H_2O$ and $RbFeCl_3·2H_2O$, *Physica B+C*, 1977, 86-88, 671-672

[25] J. A. Basten, Q. A. G. van Vlimmeren, W. J. M. de Jonge, Crystallographic and magnetic structure of $RbFeCl_3·2D_2O$ and $CsFeCl_3·2D_2O$, *Phys. Rev. B*, 1978, 18, 2179-2184.

[26] F. L. A. Machado, P. R. T. Ribeiro, J. Holanda, R. L. Rodriguez-Suarez, A. Azevedo, and S. M. Rezende, Spin-flop transition in the easy-plane antiferromagnet nickel oxide, *Phys. Rev. B*, 2017, 95, 104418

[27] Xin-Yi Wang, Lu Wang, Zhe-Ming Wang, Gang Su, and Song Gao, Coexistence of spin-Canting, Metamagnetism, and Spin-Flop in a (4,4) Layered Manganese Azide Polymer, *Chem. Mater*, 2005, 17, 6369-6380

[28] M. P. Suh, M. Y. Han, J. H. Lee, K. S. Min and C. Hyeon, One-Pot Template Synthesis and Properties of a Molecular Bowl: Dodecaaza Macrotetracycle with $\mu_3$-Oxo and $\mu_3$-Hydroxo Tricopper(II) Cores, *J. Am. Chem. Soc.*, 1998, 120, 3819;

[29] S. Meenakumari, S. Tiwary and A. Chakravarty, Synthesis, Crystal Structure, and Magnetic Properties of a Ferromagnetically Coupled Angular Trinuclear Copper(II) Complex $[Cu_3(O_2CMe)_4(bpy)_3(H_2O)](PF_6)_2$, *Inorg. Chem.*, 1994, 33, 2085

[30] T. T. Zhu, W. Sun, Y. X. Huang, Z. M. Sun, Y. M. Pan, L. Balents and J. X. Mi, Strong spin frustration from isolated triangular Cu(II) trimers in $SrCu(OH)_3Cl$ with a novel cuprate layer, *J. Mater. Chem. C*, 2014, 2, 8170

[31] N. van Well, Innovative und interdisziplinäre Kristallzüchtung, Springer Spektrum, Wiesbaden, 2016

[32] Stoe & Cie, *X-AREA*, Stoe & Cie, Darmstadt, Germany, 2002

[33] G. M. Sheldrick, A short history of SHELX, *Acta Crystallogr. Sect. A*, 2008, 64, 112−122

[34] O. Kahn, *Molecular Magnetism*, VCH-Verlag, Weinheim, 1993

[35] J. Schefer, M. Könnecke, A. Murasik, A. Czopnik, Th. Strässle, P. Keller and N. Schlumpf, Single-crystal diffraction instrument TriCS at SINQ, *Physica B*, 2000, 283-284, 168-169

[36] https://www.psi.ch/sinq/zebra/description

[37] J. Schefer, P. Fischer, H. Heer, A. Isacson, M. Koch, and R. Thut, A versatile double-axis multicounter neutron powder diffractometer, *Nucl. Instrum. Methods Phys. Res. Sect. A*, 1990, 288, 477

[38] T. Moriya, New Mechanism of Anisotropic Superexchange Interaction, *Phys. Rev.*, 1960, 120, 91-98
[39] I. Dzyaloshinsky, A thermodynamic theory of "weak" ferromagnetism of antiferromagnetics, *J. Phys. Chem. Solids*, 1958, 4, 241-255
[40] A. Weiss, H. Witte, Magnetochemie, Verlag Chemie GmbH, 1973
[41] H. Lueken, Magnetochemie, B. G. Teubner Stuttgart-Leipzig, 1990